\documentclass[12pt,a4paper]{conference}

\usepackage{fancyhdr}
\usepackage{graphicx,amsmath,amssymb,cite}
\usepackage{multind}
\makeindex{author} \makeindex{subject}

\newcommand{\beq}{\begin{equation}}
\newcommand{\eeq}[1]{\label{#1}\end{equation}}
\newcommand{\eeqn}{\end{equation}}

\newcommand{\beqa}{\begin{eqnarray}}
\newcommand{\eeqa}[1]{\label{#1}\end{eqnarray}}
\newcommand{\eeqan}{\end{eqnarray}}

\let\bar=\overbar

\newcommand{\half}{\frac{1}{2}}

\newcommand{\Dslash}{\not{\hbox{\kern-4pt $D$}}}
\newcommand{\dslash}{\not{\hbox{\kern-2pt $\del$}}}

\newcommand{\msb}{{\bar{\ssstyle M \kern -1pt S}}}

\begin{document}
\newcommand{\3}{\ss}
\newcommand{\absatz}{\vspace{2ex}\noindent}
\newcommand{\dis}{\displaystyle}
\newcommand{\txt}{\textstyle}
\newcommand{\script}{\scriptstyle}
\newcommand{\non}{\nonumber}
\newcommand{\hq}{\hspace{0.5em}}
\newcommand{\hqm}{\hspace*{-0.25em}}

\newcommand{\calO}{\mathcal{O}} \newcommand{\e}{\mathrm{e}}
\newcommand{\ii}{\mathrm{i}} \newcommand{\mpi}{\ensuremath{m_\pi}}
\newcommand{\fpi}{\ensuremath{f_\pi}}
\newcommand{\MeV}{\ensuremath{\mathrm{MeV}}}
\newcommand{\fm}{\ensuremath{\mathrm{fm}}}
\newcommand{\de}{\ensuremath{\partial}}
\newcommand{\calL}{\ensuremath{\mathcal{L}}}
\newcommand{\kv}{\ensuremath{\vec{k}}}


\Chapter{Neutron Polarisabilities from Deuteron Compton Scattering in
  $\chi$EFT} {Neutron Polarisabilities}{Harald W.~Grie\3hammer} \vspace{-6
  cm}\includegraphics[width=6 cm]{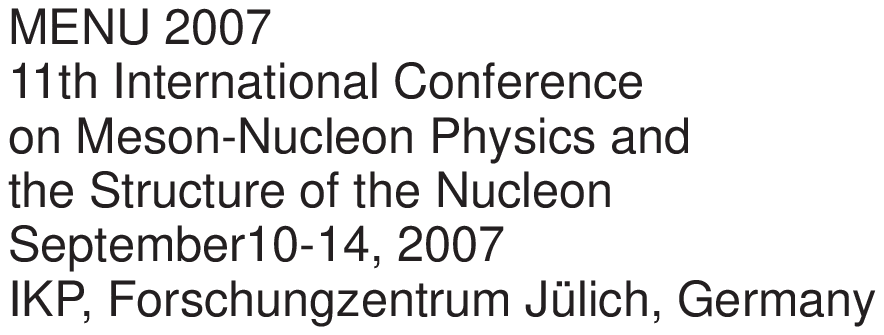}
\vspace{4 cm}

\addcontentsline{toc}{chapter}{{\it Harald W.~Grie\3hammer}} \label{authorStart}

\begin{raggedright}
{\it Harald W.~Grie\3hammer 
}\index{author}{Grie\3hammer, Harald W.}\\
Center for Nuclear Studies, Department of Physics,\\
The George Washington University, Washington DC 20052, USA\\
\bigskip\bigskip
\end{raggedright}

\index{subject}{Compton scattering}
\index{subject}{deuteron}
\index{subject}{deuteron Compton scattering}
\index{subject}{Chiral Effective Field Theory}
\index{subject}{Chiral Perturbation Theory}
\index{subject}{Polarizability (proton and neutron)}
\index{subject}{Dynamical Polarizability}
\index{subject}{Power counting}
\index{subject}{multipole expansion, for Compton scattering}
\index{subject}{}
\index{subject}{}
\index{subject}{}

\begin{center}
\textbf{Abstract}
\end{center}
Chiral Effective Field Theory is for photon energies up to $200$ MeV the tool
to accurately determine the polarisabilities of the neutron from deuteron
Compton scattering.  A multipole analysis reveals that dispersive effects from
an explicit $\Delta(1232)$ prove in particular indispensable to understand the
data at $95$ MeV measured at SAL. Simple power-counting arguments derived from
nuclear phenomenology lead to the correct Thomson limit and gauge invariance.
At next-to-leading order, the static scalar dipole polarisabilities are
extracted as identical for proton and neutron within the error-bar of
available data:
$\bar{\alpha}^n=11.6\pm1.5_{\mathrm{stat}}\pm0.6_{\mathrm{Baldin}},
\bar{\beta}^n=3.6\mp1.5_{\mathrm{stat}}\pm0.6_{\mathrm{Baldin}}$ for the
neutron, in units of $10^{-4} \;\mathrm{fm}^3$, compared to
$\bar{\alpha}^p=11.0\pm1.4_{\mathrm{stat}}\pm0.4_{\mathrm{Baldin}}$,
$\bar{\beta}^p=2.8\mp1.4_{\mathrm{stat}}\pm0.4_{\mathrm{Baldin}}$ for the
proton in the same framework. New experiments e.g.~at MAXlab (Lund) will
improve the statistical error-bar.

\section{The Problem with Neutron Polarisabilities}
\label{sec:introCompton}

As the nucleon is not a point-like spin-$\half$ target with an anomalous
magnetic moment, the photon field displaces in low-energy Compton scattering
$\gamma N\to\gamma N$ its charged constituents, inducing a non-vanishing
multipole-moment. These long-known nucleon-structure effects are for static
external fields parameterised by the electric polarisability $\bar{\alpha}$
and its magnetic counter-part $\bar{\beta}$.  For the proton, the generally
accepted static values are $\bar{\alpha}^p\approx 12\times
10^{-4}\;\fm^3,\;\bar{\beta}^p\approx 2\times 10^{-4}\;\fm^3$, with error-bars
of about $1
$.~\footnote{It is customary to measure the scalar dipole-polarisabilities in
  $10^{-4}\;\fm^3$, so that the units are dropped in the following. Notice
  that the nucleon is quite stiff.}

Does the neutron react similarly under deformations,
$\bar{\alpha}^p\approx\bar{\alpha}^n$, $\bar{\beta}^p\approx\bar{\beta}^n$?
Different types of experiments report a range of values
$\bar{\alpha}^n\in[-4;19]$: Coulomb scattering of neutrons off lead, and
deuteron Compton-scattering $\gamma d\to\gamma d$ with and without breakup,
see e.g.~\cite{polasfromdeuteron2} for a list. The latter should be a clean
way to extract the iso-scalar polarisabilities
$\bar{\alpha}^s:=\half(\bar{\alpha}^p+\bar{\alpha}^n)$ and $\bar{\beta}^s$ in
analogy to determinations of the proton polarisabilities. Experiments were
performed in Urbana at $\omega=49$ and $69$ MeV, in Saskatoon (SAL) at $94$
MeV, and in Lund (MAXlab) at $55$ and $66$ MeV. While all low-energy
extractions are consistent with small iso-vectorial polarisabilities, the SAL
data lead to conflicting analyses: The original publication~\cite{Hornidge}
gave $\bar{\alpha}^s=8.8\pm1.0$, using the Baldin sum-rule for the static
nucleon polarisabilities. Without it, Levchuk and L'vov obtained
$\bar{\alpha}^s=11\pm2,\;\bar{\beta}^s=7\pm2$~\cite{Lvov}; and Beane et
al.~found recently from all data
$\bar{\alpha}^s=13\pm4,\;\bar{\beta}^s=-2\pm3$~\cite{judith}. The extraction
being very sensitive to the polarisabilities, this seems discouraging news.

These notes outline the resolution of the puzzle and report on a new
high-accuracy determination of the nucleon polarisabilities from all Compton
scattering data. There are two main ingredients: a better understanding of
dispersive effects in the polarisabilities themselves as discussed in
Sect.~\ref{sec:dynpols}; and a model-independent determination of
meson-exchange current effects with an error-estimate,
Sect.~\ref{sec:deuteron}. As customary in proceedings, I apologise for my
biased view and refer to~\cite{polasfromdeuteron2} at least for a better list
of references.

\section{Dynamical  Polarisabilities}
\label{sec:dynpols}

The nucleon-structure effects encoded by the polarisabilities are conveniently
parameterised starting from the most general interaction between a nucleon $N$
with spin $\vec{\sigma}/2$ and an electro-magnetic field of fixed, non-zero
energy $\omega$:
\begin{eqnarray}
  \label{polsfromints}
  \lefteqn{\calL_\text{pol}=2\pi N^\dagger\big[{\alpha_{E1}(\omega)}\vec{E}^2+
  {\beta_{M1}(\omega)}\vec{B}^2+{\gamma_{E1E1}(\omega)}
  \vec{\sigma}\cdot(\vec{E}\times\dot{\vec{E}})}\\
  &&\!\!\!\!\!\!\!+{\gamma_{M1M1}(\omega)}
  \vec{\sigma}\cdot(\vec{B}\times\dot{\vec{B}})
  -2{\gamma_{M1E2}(\omega)}\sigma_iB_jE_{ij}+
  2{\gamma_{E1M2}(\omega)}\sigma_iE_jB_{ij}+\dots \big]N\non
\end{eqnarray} 
Here, the electric or magnetic (${X,Y=E,M}$) photon undergoes a transition
${Xl\to Yl^\prime}$ of definite multipolarity ${l,l^\prime=l\pm\{0,1\}}$;
${T_{ij}:=\half (\de_iT_j + \de_jT_i)}$. Its coefficients are the
\emph{energy-dependent} or \emph{dynamical polarisabilities} of the
nucleon~\cite{polas2}. Most prominently, there are six
dipole-polarisabilities. The two spin-independent ones parameterise electric
and magnetic dipole-transitions, $\alpha_{E1}(\omega)$ and
$\beta_{M1}(\omega)$.  Particularly interesting are the four
spin-polarisabilities $\gamma_{E1E1}(\omega)$, $\gamma_{M1M1}(\omega)$,
$\gamma_{E1M2}(\omega)$, $\gamma_{M1E2}(\omega)$ as they parameterise the
response of the nucleon-\emph{spin} to the photon field. Contributions from
higher ones like quadrupole polarisabilities are negligible in today's
experiments.

Polarisabilities measure the global stiffness of the nucleon's internal
degrees of freedom against displacement in an electric or magnetic field of
definite multipolarity and non-vanishing frequency $\omega$ and are identified
\emph{at fixed energy} only by their different angular dependence. Nucleon
Compton scattering provides thus a wealth of information about the internal
structure of the nucleon. 
In contradistinction to most other electro-magnetic processes,
the nucleon-structure effects in Compton scattering were however previously
not analysed in terms of a multipole-expansion at fixed energies. Instead, one
focused on the static electric and magnetic polarisabilities
$\bar{\alpha}:=\alpha_{E1}(\omega=0)$ and $\bar{\beta}:=\beta_{M1}(\omega=0)$,
which are often called ``the polarisabilities''.
While quite different frameworks could provide a consistent picture for them,
the underlying mechanisms are only properly revealed by the
energy-dependence.

Clearly, the complete set of dynamical polarisabilities does -- like in all
multipole-decompositions -- not contain more or less information about the
temporal response of the nucleonic degrees of freedom than the Compton
amplitudes. But the information is better accessible and easier to interpret,
as each mechanism leaves a characteristic signature in a particular channel.

To investigate them in a model-independent framework, we employ the unique
low-energy theory of QCD, namely Chiral Effective Field Theory $\chi$EFT. It
contains only those low-energy degrees of freedom which are observed at the
typical energy of the process, interacting in all ways allowed by the
underlying symmetries of QCD. A momentum expansion of all forces allows for
model-independent results of finite, systematically improvable accuracy and
thus for an estimate of the theoretical uncertainties encountered by
neglecting higher-order contributions.  The resulting contributions at leading
order (LO) are listed in Fig.~\ref{fig:polas} and easily
motivated~\cite{polas2}:
\begin{figure}[!htb]
\begin{center}
  \includegraphics*[width=\linewidth]{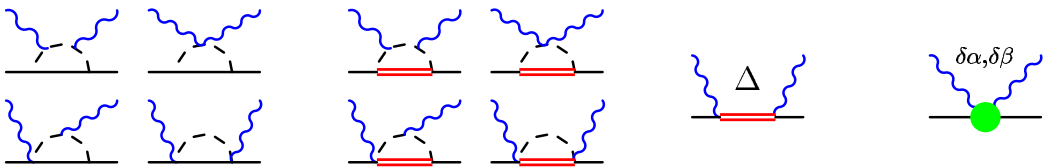}
  \caption{\label{fig:polas}The LO contributions to the nucleon
    polarisabilities. Left to right: pion cloud around the nucleon and
    $\Delta$; $\Delta$ excitations; short-distance effects.  Permutations and
    crossed diagrams not shown.  From Ref.~\cite{polas2}.}
\end{center}
\end{figure}

\textbf{(1)} Photons couple to the pions around the nucleon and around the
$\Delta$, signalled by a characteristic cusp at the one-pion production
threshold.
  
\textbf{(2)} It is well-known that the $\Delta(1232)$ as the lowest nuclear
resonance leads by the strong $\gamma N\Delta$ $M1$-transition to a
para-magnetic contribution to the static magnetic dipole-polarisability
$\bar{\beta}_{\Delta} =+[7\dots13]$ and a characteristic resonance-shape,
cf.~the Lorentz-Drude model of classical electrodynamics.
  
\textbf{(3)} As the observed static value $\bar{\beta}^p\approx 2$ is smaller
by a factor of $5$ than the $\Delta$ contribution, a strong dia-magnetic
component must exist. The resultant fine-tuning at zero photon-energy is
unlikely to hold once the evolution with the photon energy is considered: If
dia- and para-magnetism are of different origin, they involve different scales
and hence different energy-dependences. We sub-sume this short-distance
Physics which is at this order not generated by the pion or $\Delta$ into two
\emph{energy-independent} low-energy coefficients
$\delta\alpha,\;\delta\beta$.

The cornucopia of Compton data on the proton below $200\;\MeV$ determines
these to be indeed anomalously large, $\delta\alpha=-5.9\pm1.4,\;
\delta\beta=-10.7\pm1.2$, justifying their inclusion at leading order. As
expected, $\delta\beta$ is dia-magnetic. The resulting static proton
polarisabilities
\begin{equation}
  \label{eq:protonvalues1}
  \dis\bar{\alpha}^p=11.0\pm1.4_\mathrm{stat}\pm0.4_\text{Baldin}\;\;\;\;
  ,\;\;\;\;\bar{\beta}^p=2.8\mp1.4_\mathrm{stat}\pm0.4_\text{Baldin}
\end{equation}
compare both in magnitude and uncertainty favourably with other
state-of-the-art results~\cite{polas2}. Higher-order corrections contribute
an error of about $\pm1$ not displayed here as the statistics dominates the
total error.

\begin{figure}[!htb]
\begin{center}
  \includegraphics[width=0.325\linewidth]{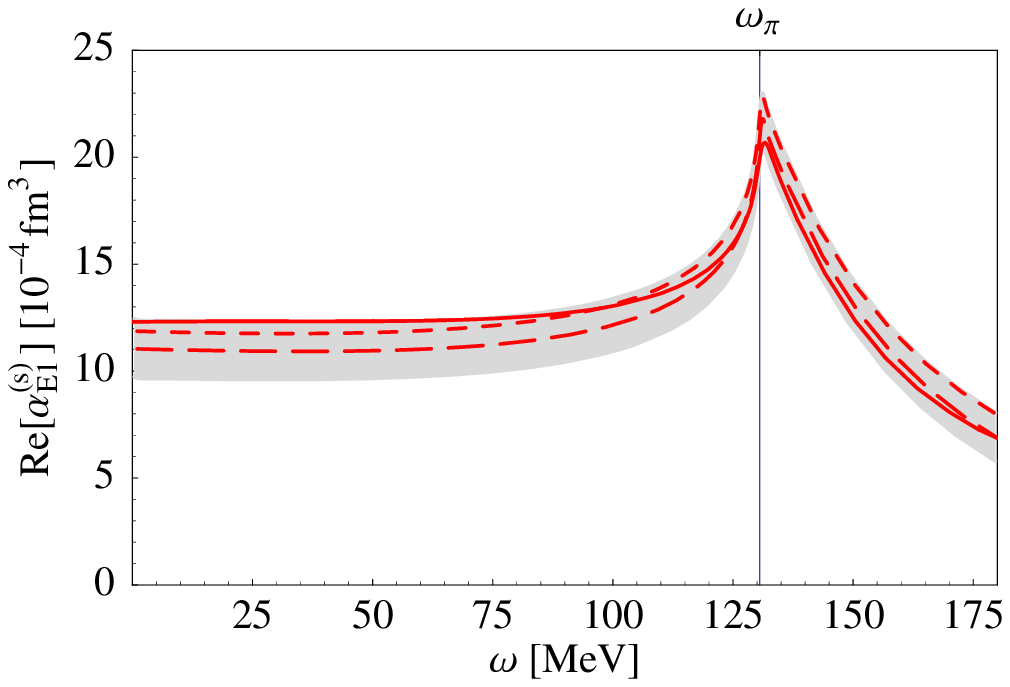}\hfill
  \includegraphics[width=0.325\linewidth]{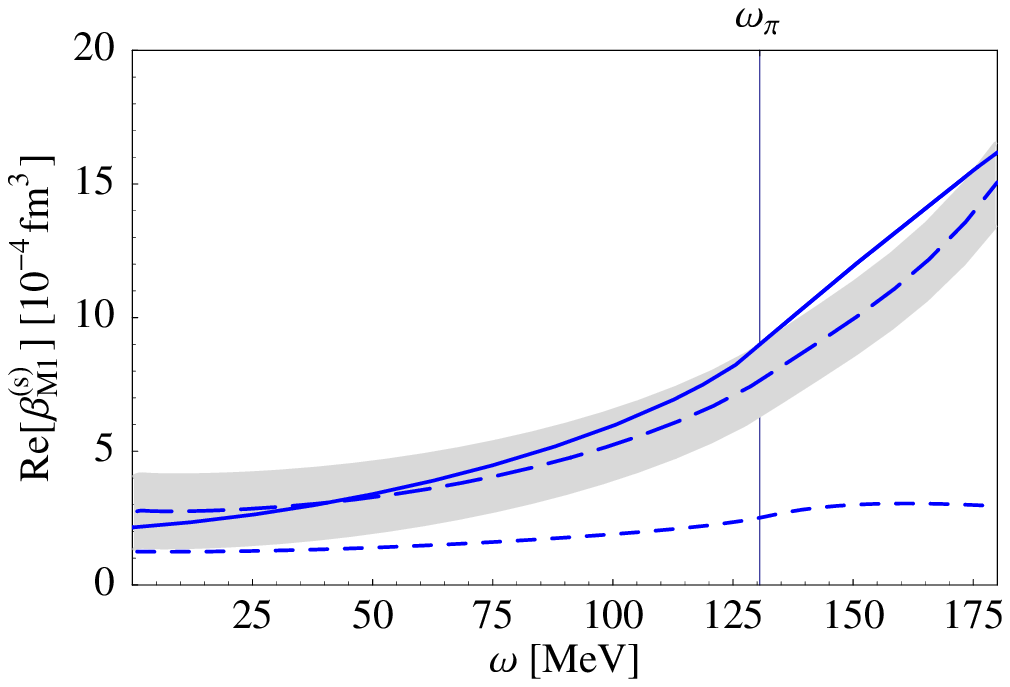}\hfill
  \includegraphics[width=0.325\linewidth]{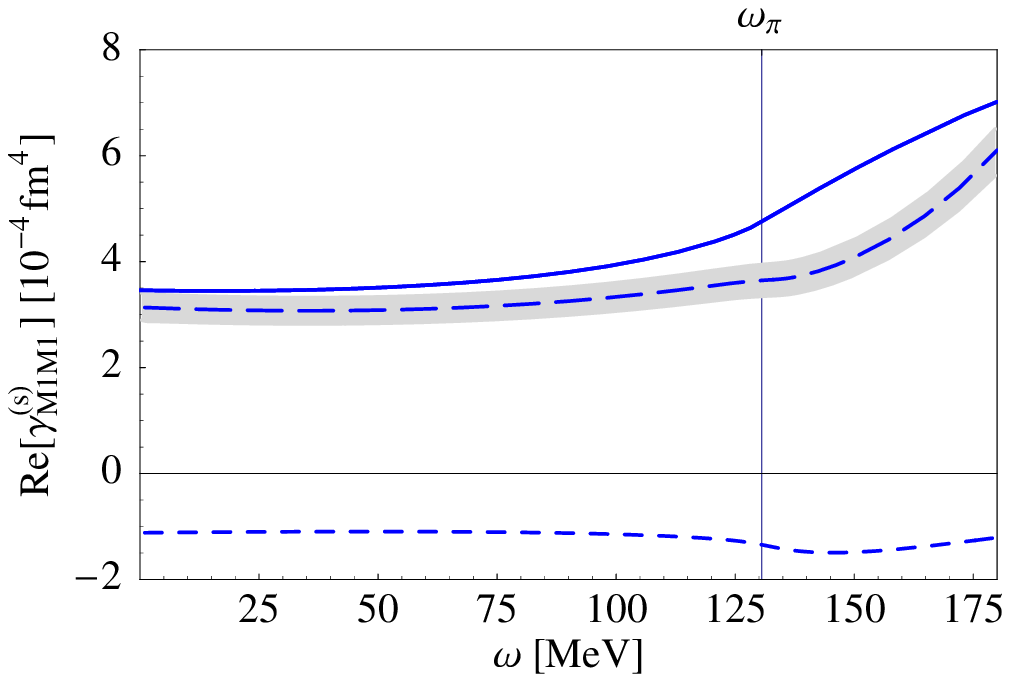}
  \caption{\label{fig:dipolepols} Spin-independent (left, right) and example
    of spin-dependent (right) dipole-polarisabilities, predicted by Dispersion
    Theory (solid) and $\chi$EFT with (dashed; band: fit-errors) and without
    (dotted) explicit $\Delta$. $\omega_\pi$: one-pion production threshold.
    From Ref.~\cite{polas2}.}
\end{center}
\end{figure}

\absatz With the parameters now fixed, the energy-dependence of all
polarisabilities is fixed. $\chi$EFT predicts them at LO to be identical for
the proton and neutron. We will confirm this in Sect.~\ref{sec:deuteron}.  The
dipole-polarisabilities show the expected behaviour. No low-energy degrees of
freedom inside the nucleon are missing.  Dispersion is large for $\omega\in
[80;200]\;\mathrm{MeV}$ where most experiments to determine polarisabilities
are performed.  Most notably even well below the pion-production threshold is
the strong energy-dependence induced into $\beta_{M1}(\omega)$ and all
polarisabilities containing an $M1$ photon by the unique signature of the
$\Delta$-resonance: Truncating the Taylor-expansion at order $\omega^2$
under-estimates $\beta_{M1}(\omega=95\;\MeV)-\bar{\beta}\approx
1.7$~\cite{Lvov}, while the multipole-analysis gives $\approx 4$
(Fig.~\ref{fig:dipolepols}). The traditional approximation of
$\beta_{M1}(\omega)$ as ``static-plus-small-slope'',
$\bar{\beta}+\omega^2\bar{\beta}_\nu$, is inadequate.  Not surprisingly, this
contribution is most pronounced at large momentum-transfers, i.e.~backward
angles, and thus is the major source of confusion in deuteron Compton
scattering, as Fig.~\ref{fig:results} will show. Figure~\ref{fig:dipolepols}
reveals the good agreement between the measured value of $\bar{\beta}^p$ and
the prediction in $\chi$EFT without explicit $\Delta$ as accidental: The pion
is not dispersive enough to explain the energy-dependence of $\beta_{M1}$.

That the two short-distance parameters $\delta\alpha,\;\delta\beta$ suffice to
describe the data up to $\omega\approx200\;\MeV$~\cite{polas2} leads to three
constraints on their explanation:

\textbf{(1)} Like $\delta\alpha,\;\delta\beta$, the effect must be
$\omega$-independent over a wide range.
  
\textbf{(2)} Albeit it must lead to the values for $\delta\alpha,\;\delta\beta$
predicted in $\chi$EFT, it must be absent at least in the pure
spin-polarisabilities $\gamma_{E1E1},\;\gamma_{M1M1}$.
  
\textbf{(3)} Its prediction for the proton and neutron must be similar because
iso-vectorial effects are small and
energy-independent~\cite{deuteronpaper,polas2}.

\section{Embedding the Nucleon in the Deuteron}
\label{sec:deuteron}
\index{subject}{Power counting}

Since free neutrons can often not be used in experiments, their properties are
usually extracted from data taken on few-nucleon systems by dis-entangling
nuclear-binding effects. $\chi$EFT allows to subtract two-body contributions
from meson-exchange currents and wave-function dependence from data with
minimal theoretical prejudice and an estimate of the theoretical
uncertainties.

A consistent description must also give the correct Thomson limit, i.e.~the
exact low-energy theorem which is a consequence of gauge
invariance~\cite{Friar,Arenhoevel}.
\begin{figure}[!htbp]
  \begin{center}
    \includegraphics*[width=0.8\linewidth]{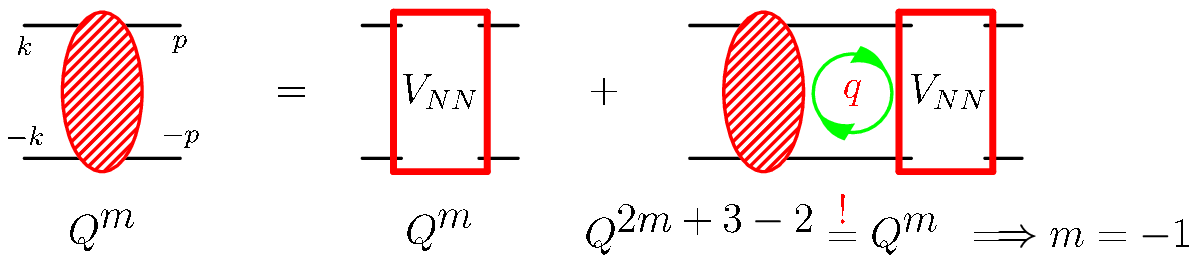}
  \caption{\label{fig:consistency}
    On the consistency of $NN$ power-counting in $\chi$EFT. From
    Ref.~\cite{hg}.}
\end{center}
\end{figure}
Its verification is straight-forward in the 1-nucleon sector, where the
amplitude is perturbative. In contradistinction, the two-nucleon amplitude
must be non-perturbative to accommodate the shallow bound-state: All terms in
the LO Lippmann-Schwinger equation of $NN$-scattering,
Fig.~\ref{fig:consistency}, including the potential, must be of the same order
when all nucleons are close to their non-relativistic mass-shell.  Otherwise,
one of them could be treated as perturbation of the others and a low-lying
bound-state would be absent.  Picking the nucleon-pole in the
energy-integration $E\sim\frac{\kv^2}{2M}$ leads therefore to the consistency
condition that the $NN$-scattering amplitude $T_{NN}$ must be of order
$Q^{-1}$, irrespective of the potential used. $Q$ is a typical low-momentum
scale of the process under consideration, e.g.~the inverse $\mathrm{S}$-wave
scattering length. It does therefore not suffice to determine the relative
strength of forces and potentials in $\chi$EFT just by counting the number of
momenta. This has long been recognised in ``pion-less'' EFT, but is only an
emerging communal wisdom in the chiral version~\cite{hg,NoggaBirse,hgpc}. 

\begin{figure}[!htbp]
  \begin{center}
    \includegraphics*[width=\linewidth]{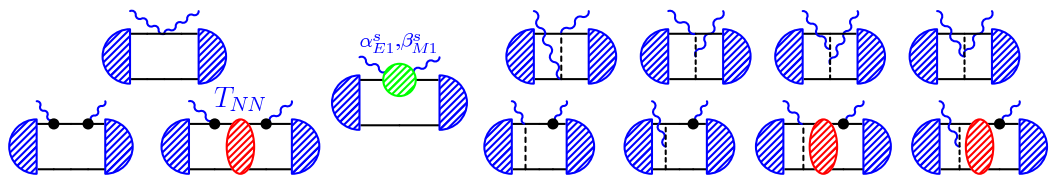}
  \caption{\label{fig:dgraphs}
    Deuteron Compton scattering in $\chi$EFT to NLO. Left: one-body part (dot:
    electric/magnetic coupling; blob: nucleon polarisabilities of
    Fig.~\ref{fig:polas}). Right: two-body part (pion-exchange currents).
    Permutations and crossed graphs not shown.  From
    Ref.~\cite{polasfromdeuteron2}.}
\end{center}
\end{figure}

In deuteron Compton scattering, this mandates inclusion of $T_{NN}$ for all
graphs in which both nucleons propagate close to their mass-shell between
photon absorption and emission, i.e.~in which the photon energy
$\omega\lesssim 50\;\MeV$ does not suffice to knock a nucleon far off its
mass-shell~\cite{polasfromdeuteron2,hg}. Figure~\ref{fig:dgraphs} lists the
contributions to Compton scattering off the deuteron to next-to-leading order
NLO in $\chi$EFT. At higher photon energies $\omega\gtrsim 60\;\MeV$, on can
show that the nucleon is kicked far off its mass-shell, $E\sim|\kv|$, and the
amplitude becomes perturbative. This is intuitively clear, as the nucleon has
only a very short time ($\sim1/\omega$) to scatter with its partner before the
second photon has to be radiated to restore the coherent final state. The
diagrams which contain $T_{NN}$ in Fig.~\ref{fig:dgraphs} are therefore less
important for larger $\omega$, together with some of the other diagrams.
Indeed, the nucleon propagator scales then as $1/Q\sim1/\omega$ and thus
becomes static, with each re-scattering process in $T_{NN}$ suppressed by an
additional power of $Q$.

\begin{figure}[!htb]
 \includegraphics*[width=0.31\linewidth]{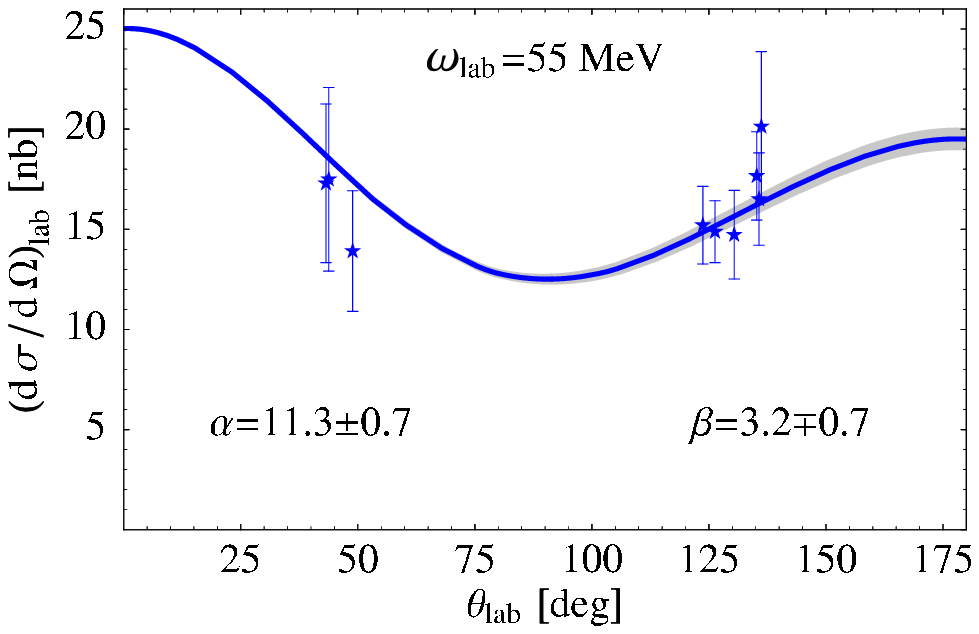}
 \hfill
 \includegraphics*[width=0.31\linewidth]{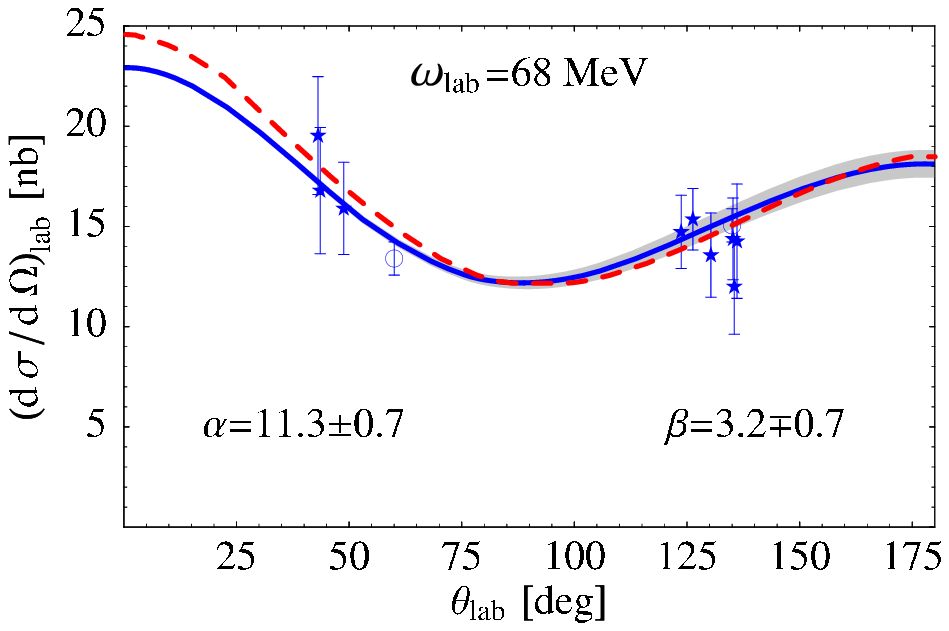}
 \hfill
 \includegraphics*[width=0.32\linewidth]{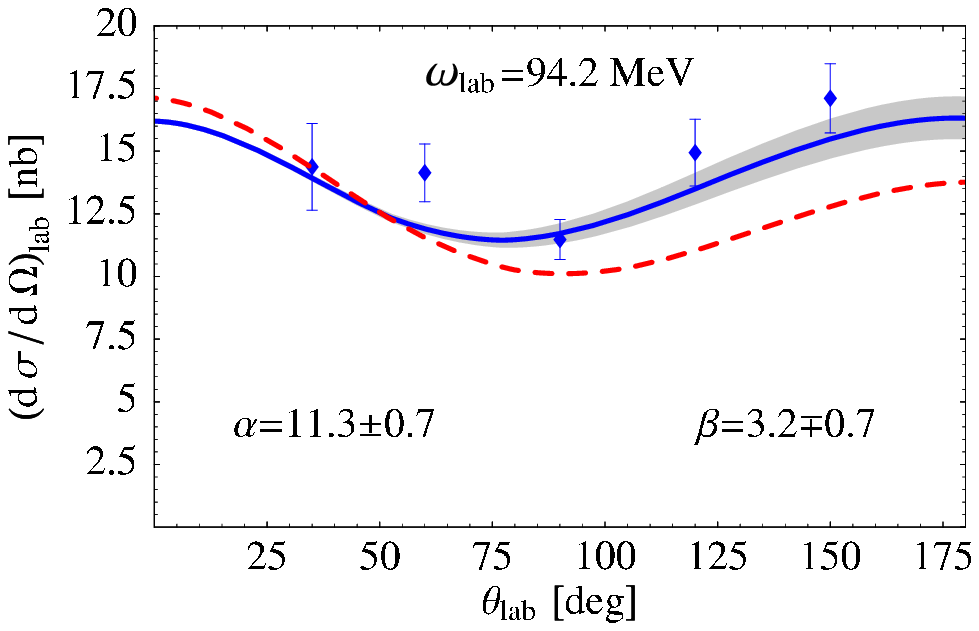}

 \caption{Examples for the 1-parameter fit result using the Baldin sum rule
   (solid, with stat.~uncertainty), compared to $\chi$EFT without explicit
   $\Delta(1232)$ ($\calO(p^3)$, dashed).  From
   Ref.~\cite{polasfromdeuteron2}.}
\label{fig:results}
\end{figure}

\absatz We implemented rescattering by the Green's function method described
in~\cite{Karakowski,Lvov,polasfromdeuteron2}. The calculation is
parameter-free when the short-distance coefficients
$\delta\alpha,\;\delta\beta$ are taken over from the proton -- as justified by
the $\chi$EFT prediction that iso-vectorial contributions are suppressed by
one order. The nucleon- and nuclear-structure contributions separate at this
(and the next) order. While the two-nucleon piece does not contain the
$\Delta(1232)$-resonance in the intermediate state at this order as the
deuteron is an iso-scalar target, this does not hold for the polarisabilities,
as seen in Sect~\ref{sec:dynpols}. Figure~\ref{fig:results} also shows that
the strong energy-dependence from the $\Delta$ is indeed pivotal to reproduce
both shape and normalisation of the $94\;\MeV$ data in particular at
back-angles without significantly changing the static polarisabilities, but is
negligible at lower energies. Thus, we argue that the discrepancy between the
SAL data and experiments at lower energies is
resolved~\cite{polasfromdeuteron2,deuteronpaper}.

\begin{figure}
 \includegraphics*[width=0.31\linewidth]{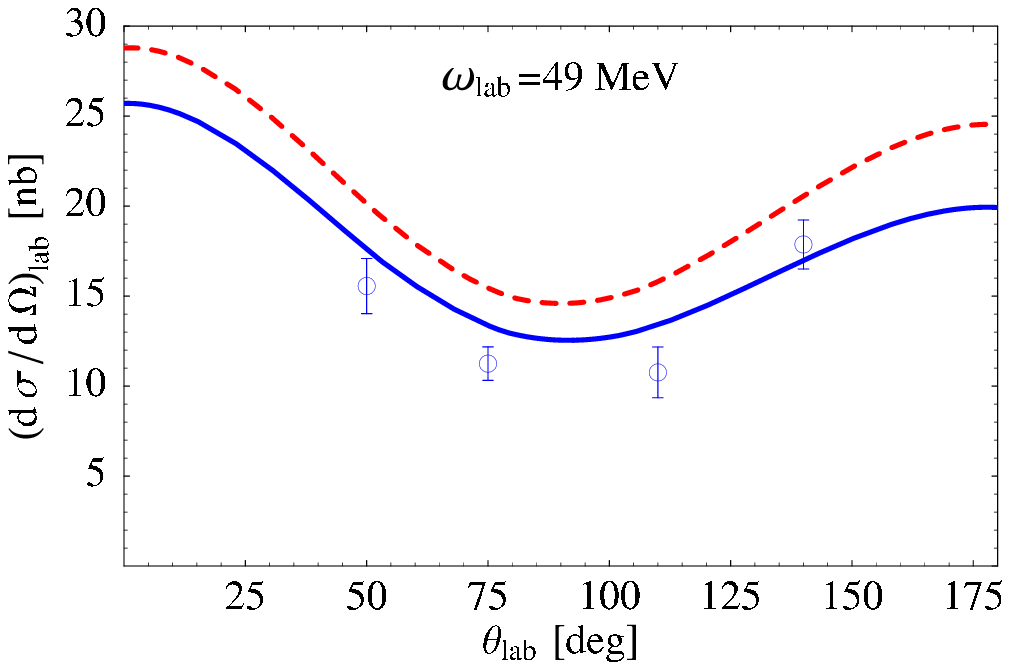}
 \hfill
 \includegraphics*[width=0.31\linewidth]{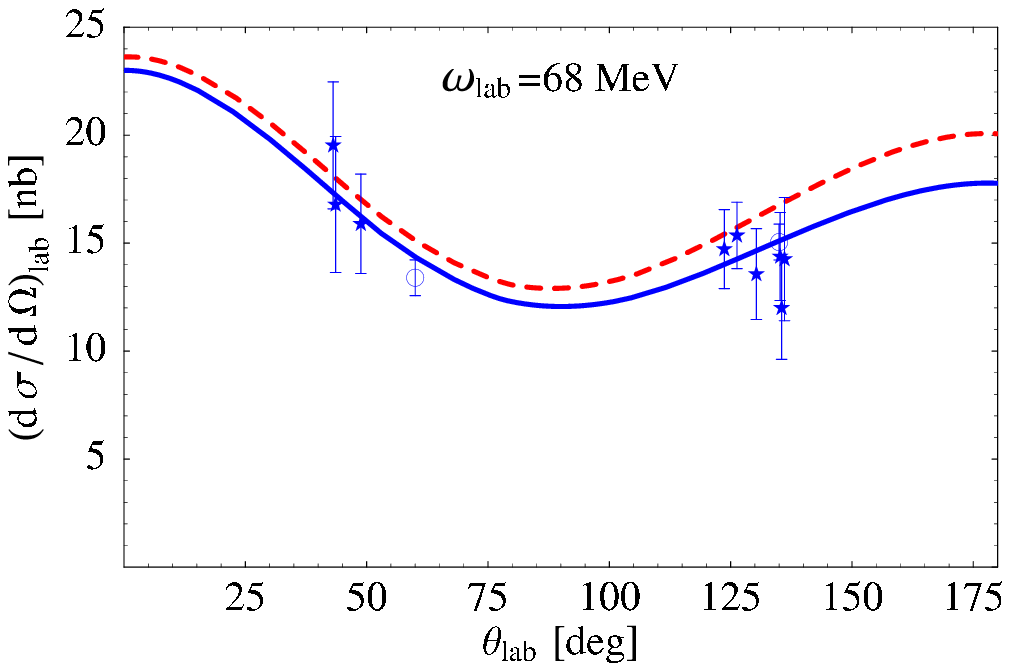}
 \hfill
 \includegraphics*[width=0.32\linewidth]{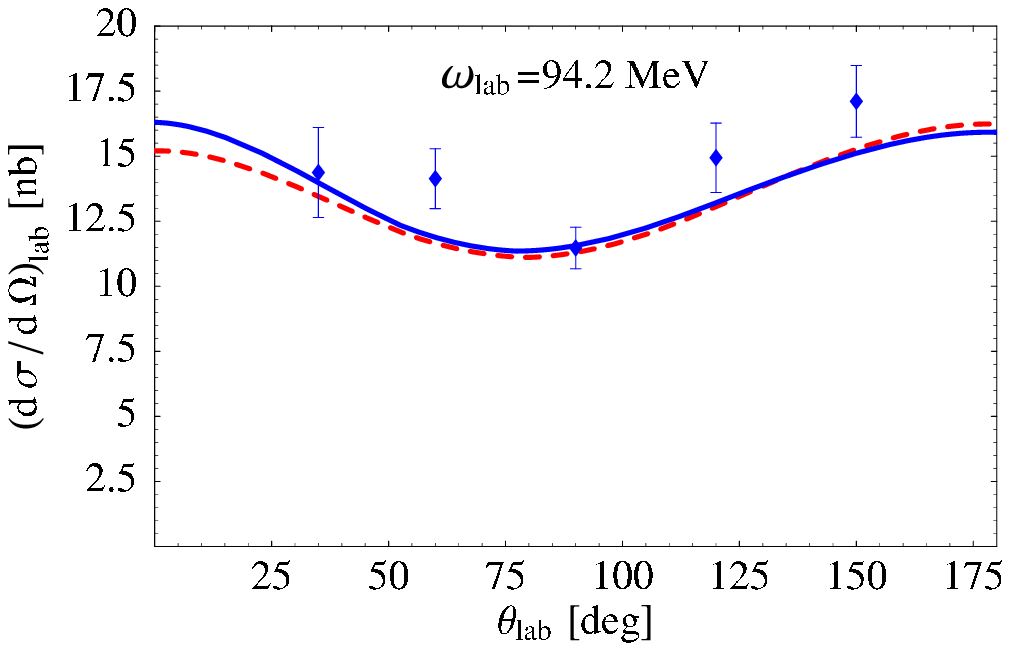}
 \caption{Examples of prediction using proton polarisabilities with (solid)
   and without (dashed) $NN$-rescattering in intermediate states. From
   Ref.~\cite{polasfromdeuteron2}.}
\label{fig:1}
\end{figure}

The power-counting at the heart of $\chi$EFT implies several cross-checks:
First, it must automatically reproduce the Thomson limit as an exact LO
result, with all corrections cancelling order by order as
$\omega\to0$~\cite{Friar}.  Fortunately, Arenh\"ovel showed long before
$\chi$EFT was formulated that it is indeed exactly recovered from the diagrams
which $\chi$EFT classifies as LO at low energies, and that all diagrams which
couple photons to meson-exchange currents sum up to zero at zero
energy~\cite{Arenhoevel}. The numerical calculation confirms
this~\cite{polasfromdeuteron2}.

Secondly, $\chi$EFT demotes at higher photon energies all graphs with $T_{NN}$
in the intermediate state to higher orders. The difference to the previous
$\chi$EFT calculations~\cite{deuteronpaper,judith} which were
tailored to high photon energies should therefore decrease with increasing
$\omega$. This is indeed found, see Fig.~\ref{fig:1}.

\begin{figure}
 \includegraphics*[width=0.31\linewidth]{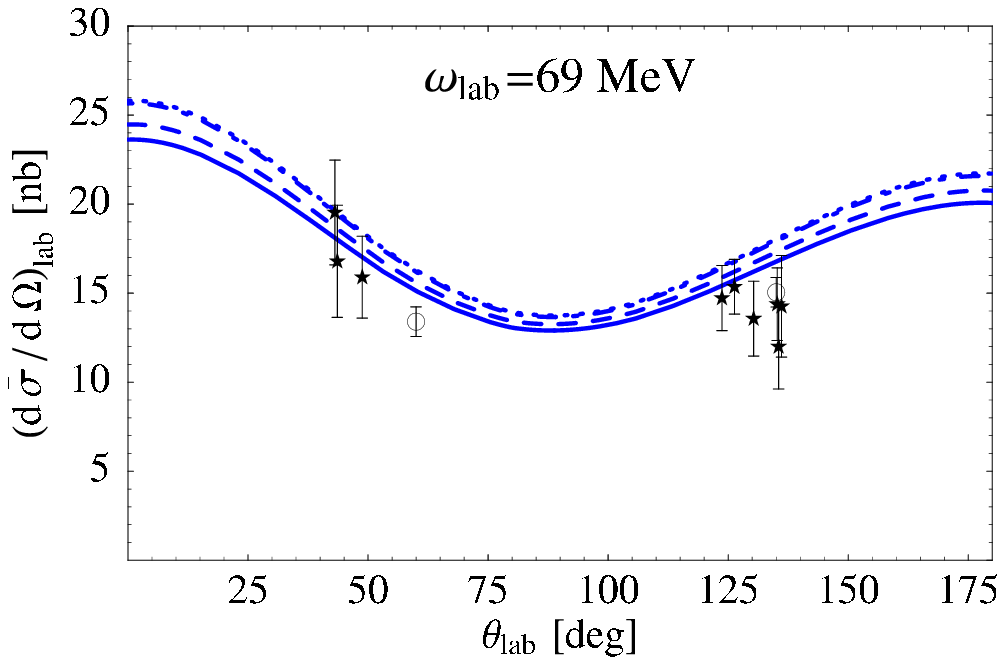}
 \hq
 \includegraphics*[width=0.31\linewidth]{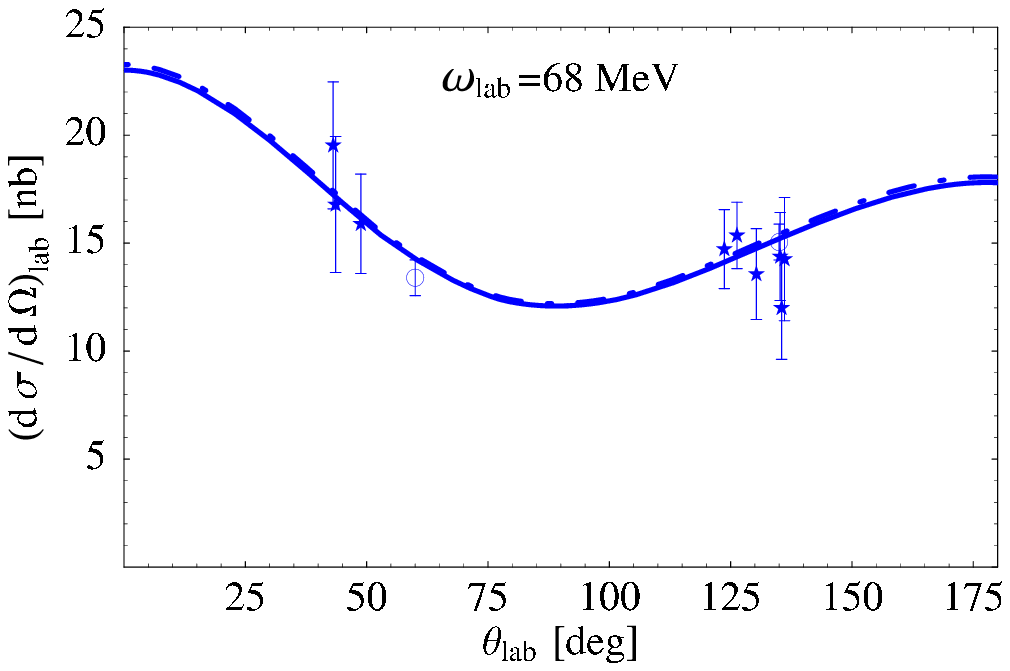}
 \hfill
 \includegraphics*[width=0.32\linewidth]{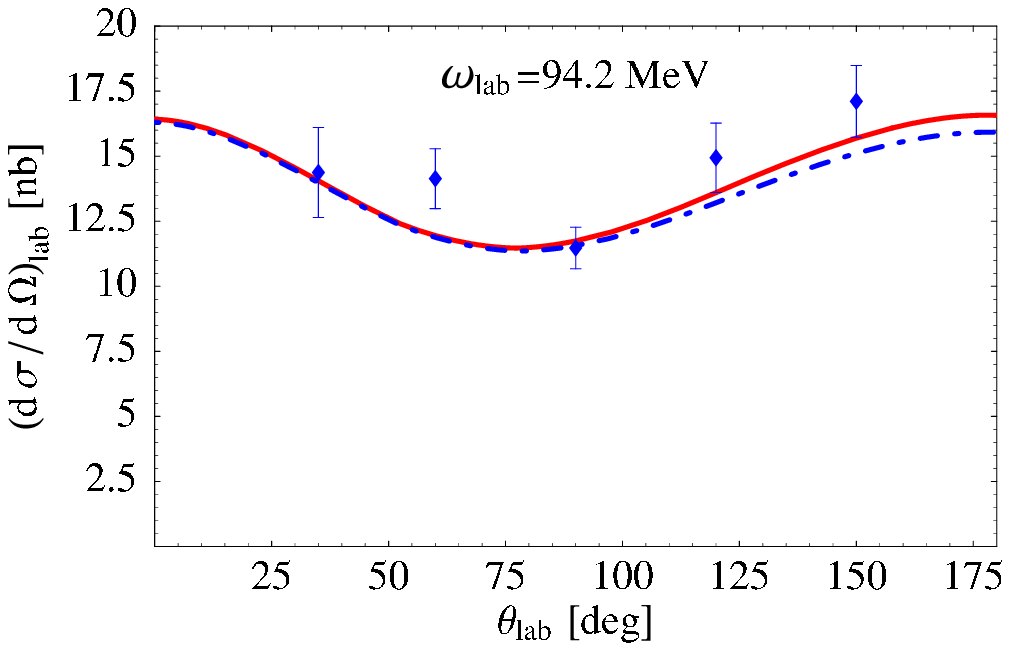}

 \caption{Examples of dependence on higher-order effects, with iso-scalar
   polarisabilities at proton values. Left (without $T_{NN}$) and centre (with
   $T_{NN}$): different deuteron wave-functions (solid: NNLO $\chi$EFT;
   dot-dashed: AV18; dashed: CD-Bonn; dotted: Nijmegen 93). Right: Dependence
   on the $NN$-potential for $T_{NN}$ (solid: LO $\chi$EFT; dot-dashed: AV18).
   From Ref.~\cite{polasfromdeuteron2}.}
\label{fig:reductions}
\end{figure}

Another consequence is a substantially reduced dependence on the deuteron
wave-function, see Fig.~\ref{fig:reductions}. With the long-range part fixed
by the deuteron binding energy and one-pion exchange, different wave-functions
and potentials correspond to different assumptions about the short-distance
dynamics of $NN$-scattering. Different answers would hence indict
model-dependence, i.e.~sensitivity to details of Physics at scales where a
description in terms of the low-energy degrees of freedom breaks down. In the
model-\emph{independent} approach of $\chi$EFT, answers from different
potentials and wave-functions agree within the theoretical accuracy,
i.e.~serve to estimate higher-order contributions. While the Thomson limit is
universal for $\omega\to0$~\cite{Friar,Arenhoevel}, the dependence on the
deuteron wave-function used is now also at higher energies virtually
eliminated compared to previous approaches~\cite{deuteronpaper,judith}.

Figure ~\ref{fig:reductions} shows that the result is also quite insensitive
to the potential from which $T_{NN}$ is found. The maximal difference is
smaller than $3$\% between constructing it from AV18 and a one-pion exchange
with a crude parameterisation of short-distance effects as two point-like,
momentum-independent contact operators. In $\chi$EFT, these differences come
from $NN$ interactions which are suppressed by $Q^2\approx(1/7)^2$, in line
with the spread found.

\absatz Finally, we test whether the neutron and proton polarisabilities are
indeed similar by fitting the two short-distance parameters
$\delta\alpha,\;\delta\beta$ to all deuteron Compton scattering data below
$100\;\MeV$~\cite{polasfromdeuteron2}. The iso-scalar Baldin sum rule
$\bar{\alpha}^s+\bar{\beta}^s=14.5\pm0.6$ is in excellent agreement with our
2-parameter fit, serving in the next step as input to model-independently
determine the iso-scalar, spin-independent dipole polarisabilities of the
nucleon at zero energy:
\begin{equation}
  \label{eq:neutronpols}
  \bar{\alpha}^s=11.3\pm0.7_\mathrm{stat}\pm0.6_\mathrm{Baldin}\pm1_\mathrm{th}
  \;,\; 
  \bar{\beta}^s =3.2\mp0.7_\mathrm{stat}\pm0.6_\mathrm{Baldin}\pm1_\mathrm{th}
\end{equation}
We estimate the theoretical error to be $\pm1$ from typical higher-order
contributions in the one- and two-nucleon sector. Comparing this with our
analysis \eqref{eq:protonvalues1} of all proton Compton data below $170\;\MeV$
by the same method, we conclude that the proton and neutron polarisabilities
are to this leading order identical within (predominantly statistical) errors
and confirm the $\chi$EFT prediction. In particular, the proton and neutron
show only a small but very similar deformation when put between the poles of a
magnet: $\bar{\beta}^p\approx\bar{\beta}^n\approx 3$.

\section{Concluding Questions}
\label{sec:perspectives}

Dynamical polarisabilities test the global response of the nucleon to the
electric and magnetic fields of a real photon with non-zero energy and
definite multipolarity. They answer the question which internal degrees of
freedom govern the structure of the nucleon at low energies and are defined by
a multipole-expansion of the Compton amplitudes. While they do not contain
more or less information than the corresponding Compton scattering amplitudes,
the facts are more readily accessible and easier to interpret. Dispersive
effects in particular from the $\Delta(1232)$ are necessary to accurately
extract the static polarisabilities of the nucleon from all data. Future work
includes:

(i) The non-zero width of the $\Delta$ and higher-order effects from the
pion-cloud become crucial in the resonance region.
  
(ii) A multipole-analysis of Compton scattering at fixed energies from
doubly-polarised, high-accuracy experiments provides a new avenue to extract
the energy-dependence of the six dipole-polarisabilities per nucleon, both
spin-independent and spin-dependent~\cite{polas2}. This will in particular
further our knowledge on the spin-polarisabilities which characterise the
spin-structure of the nucleon. A concerted effort of planned and approved
experiments at $\omega\lesssim300\;\MeV$ is under way: polarised photons on
polarised protons, deuterons and ${}^3$He at TUNL/HI$\gamma$S; tagged protons
at S-DALINAC; polarised photons on polarised protons at MAMI. An unpolarised,
running experiment on the deuteron at MAXlab covers a wide range of energies
and angles. With at present only 29 (un-polarised) points for the deuteron in
a small energy range of $\omega\in[49;94]$ MeV and error-bars on the order of
$15\%$, these high-quality data will provide better information on the neutron
polarisabilities and allow one to zoom in on the proton-neutron differences.

(iii) Choudhury et al.~found that Compton scattering on ${}^3$He also shows
high sensitivity to the neutron polarisabilities~\cite{Choudhury:2007bh}. In a
coordinated effort, we now investigate which observables in proton, deuteron
and ${}^3$He Compton scattering are most sensitive to combinations of
polarisabilities in $\chi$EFT. Of particular interest are polarisation
asymmetries because of their sensitivity to the experimentally practically
undetermined dipole spin-polarisabilities.

\absatz Enlightening insight into the electro-magnetic structure of the
nucleon has already been gained from combining Compton scattering off nucleons
and few-nucleon systems with $\chi$EFT and the (energy-dependent) dynamical
polarisabilities; and a host of activities should add to it in the coming
years.

\section*{Acknowledgements}
I am grateful for financial support by the National Science Foundation (CAREER
grant PHY-0645498), US Department of Energy (DE-FG02-95ER-40907) and Deutsche
Forschungsgemeinschaft (GR1887/3-1).  Foremost, I thank my collaborators --
R.P.~Hildebrandt, T.R.~Hemmert, B.~Pasquini and D.R.~Phillips -- for a lot of
fun!

\end{document}